\documentclass[conference]{IEEEtran}

\DeclareUnicodeCharacter{2212}{-}
\DeclareUnicodeCharacter{25CF}{$\bullet$}
\usepackage[T1]{fontenc}
\usepackage[utf8]{inputenc} 
\usepackage[T1]{fontenc}    
\usepackage{hyperref}
\usepackage{soul}
\usepackage{xcolor}
\usepackage{url}            
\usepackage{booktabs}       
\usepackage{amsfonts}       
\usepackage{microtype}      
\usepackage{lipsum}		
\usepackage{amsmath}
\usepackage{mathtools}
\usepackage{graphicx}
\usepackage{subfig} 
\usepackage[ruled,vlined]{algorithm2e}
\usepackage{url}
\usepackage{pdfpages}
\usepackage{float}
\usepackage[utf8]{inputenc}
\usepackage{color} 
\usepackage{soul} 
\usepackage{multirow}
\ifCLASSINFOpdf
\else
\fi
\usepackage{amsmath}
\usepackage[cmintegrals]{newtxmath}
\def\BibTeX{{\rm B\kern-.05em{\sc i\kern-.025em b}\kern-.08em
    T\kern-.1667em\lower.7ex\hbox{E}\kern-.125emX}}
\begin{document}


\title{Security Considerations for Virtual Reality Systems}
\author{\IEEEauthorblockN{ Karthik Viswanathan\IEEEauthorrefmark{1}  }
\IEEEauthorblockA{\textit{\IEEEauthorrefmark{1}Department of Software Engineering and Game Development} \\
\textit{Kennesaw State University, GA, USA  }\\
 kviswana@students.kennesaw.edu}
\and
\IEEEauthorblockN{ Abbas Yazdinejad\IEEEauthorrefmark{2}}
\IEEEauthorblockA{\textit{\IEEEauthorrefmark{2}Cyber Science Lab, School of Computer Science} \\
\textit{University of Guelph, Ontario, Canada }\\
ayazdine@uoguelph.ca}

}

\maketitle
\begin{abstract}
There is a growing need for authentication methodology in virtual reality applications. Current systems assume that the immersive experience technology is a collection of peripheral devices connected to a personal computer or mobile device. Hence there is a complete reliance on the computing device with traditional authentication mechanisms to handle the authentication and authorization decisions. Using the virtual reality controllers and headset poses a different set of challenges as it is subject to unauthorized observation, unannounced to the user given the fact that the headset completely covers the field of vision in order to provide an immersive experience. As the need for virtual reality experiences in the commercial world increases, there is a need to provide other alternative mechanisms for secure authentication. In this paper, we analyze a few proposed authentication systems and reached a conclusion that a multi-dimensional approach to authentication is needed to address the granular nature of authentication and authorization needs of a commercial virtual reality applications in the commercial world.
\end{abstract}
\section{INTRODUCTION}\label{s1}
In order to evaluate proposed security solutions for commercial virtual reality applications  \cite{v1, v2}, we identified two possible virtual reality applications. These were selected based on solving different problems within very different industries. We identified potential security requirements by analyzing the use cases. Authentication and authorization solutions \cite{a1,a2} for virtual reality systems from the literature were reviewed against these requirements. While each of the systems proposed had their strengths, we concluded that none of them met the requirements that we identified by themselves. Therefore, we concluded that in reality, the commercial VR application will require a high level of granularity. In order to accomplish this, authentication needs to be more than just verifying the identity of the user but has to be a multi-dimensional real time process. 

Most virtual reality systems today are coupled directly to a PC \cite{v1}.  It is easy to assume that the application that provides the immersive experience runs on the PC using the VR headset and controllers as peripherals. Security in essence is the ability to make two specific decisions – authentication and authorizations. As commercial applications of VR become commonplace, some of these basic assumptions will change. The VR system may no longer be tethered to or hosted on a PC with a keyboard. Therefore, the keyboard based or 2D image-based authentication will not be feasible. Authentication poses a different challenge in that the headset typically covers the eye and blocks out the view of observers, cameras that surround the user in the physical world. This opens up the possibility for a keen observer to notice the motions of the controllers and therefore know the authentication credentials. 

Security involves making authentication and authorization decisions \cite{P2019}. For most part, authentication is the process of discerning the identity of the user. Once that identity is established, authorization involves making decisions on what resources this identified user has access to. 

In a commercial VR application, especially in an environment where the user may move to different locations through the VR experience, it is easy to envision that the authorization decisions will take into consideration the geolocation of the user and require additional special authentication to access certain specific resources. This adds a requirement that the authentication process cannot be limited to a one time gating event at the point of entry, but one that continuously updates the authentication object in multiple dimensions.

In this context, we concluded that authentication is no longer a means to determining identity but has to be a multidimensional validation that allows one to incorporate levels of complexity in the authentication and take environmental factors into account. Authorization decisions will consume this multidimensional identity to make authorization decisions. 

\section{AUTHENTICATION MECHANISMS COMPARED}\label{s2}
\subsection{Knowledge Driven Biometric authentication}
This methodology involves the application of known human behaviors when conducting the process of biometric authentication. An authentication scheme called RubikBiom was developed based on Guiard’s kinematic chain model. Factors such as the pose of the non-dominant hand, the motion of the dominant hand and other such factors are taken together to come up with the authentication of the user.  This involves complex algorithms that measure behavioral patterns and other such knowledge regarding the user. This involves deep learning technologies applied to human behavior. In a limited study, they were able to demonstrate 98.91\% accuracy in authentication.

\subsection{RubikAuth}
Another experimental authentication software is the RubikAuth. Taking its name from the inventor of the iconic Rubik’s Cube, the RubikAuth essentially acts as a 9-digit pin input overlayed over each face of a multicolored cube but the rear. Users are instructed to dial in their authentication code by selecting a number and pulling the trigger on the non-cube hand. Using a range of selection methods like eye tracking or head direction, tests were able to conclude with a 98.52\% failure rate of “attack” \cite{8}. This data shows that unique VR systems authentication systems can be created in such a way that attackers in the real-world struggle to replicate the correct solution \cite{6}.
\subsection{OcuLock}
Human Visual System (HVS) includes not just the eyeball, but eyelids, nerves around the eye, extraocular muscles, cells etc. OcuLock therefore applies this complex algorithm that takes into account the various aspects of HVS by stimulating the eye at predefined times to activate the HVS behavior. Error rates vary between 3.55\% and 4.97\%. 
\subsection{BioMove}
An international group produced a biometric identification process known as BioMove, which prompts the user to perform certain simple tasks while reading their motion. Based on the user’s unique manner of motion, the software then verifies their identity. Surprisingly, the software managed to predict the motion of existing participants over 95\% of the time, though the results of attempted penetration testing fluctuated more.

On the lower end, most subjects trying to mimic another’s movements reported low success, but those with similar physical structure were able to succeed up to half the time \cite{9} Because of this, the software would not be considerably safe as the only method of security but might serve as a good preliminary or secondary test, especially with some refinement.

Furthermore, the combination of technologies such as the RubikAuth and BioMove could result in much higher security while keeping the time of authentication entry under 5 seconds. Another option would simply be to implement existing biometric verification into the hardware. Controllers could simply have fingerprint scanners mounted, or eye trackers could be augmented with retinal scans. Due to the additional price, however, these types of securities will likely be reserved for industrial usage rather than consumer use.

\section{USE CASES CONSIDERED – METHODOLOGY}\label{S3}
In order to study the need to a granular approach to authentication and authorization, two potential use cases were studied. Both these were looked at, not from the perspective of what is currently available, but more from the perspective of where the industry is headed. Two unrelated industries were selected and reviewed to understand the need for granular real-time authentication and authorization. The multidimensional security model was considered in light of the two use cases.
\subsection{VR application to solve pick problem in warehouse}
In a warehouse, there has always been a challenge to optimize the ability for the pickers to navigate and pick specific products from specific locations. This age old problem has been tackled in several ways, including the documentation of the locations of various products in the warehouse. Typically a warehouse management system shows the entire inventory and the locations. The reality is that things get moved and are not always tracked to a great extent. 

We can envision the pickers wearing a VR headset and being able to see the location of the item that they need to pick and navigate the warehouse in an efficient way. With the VR headset, the pickers do not have the view of the real world and are navigating based of the virtual reality being presented to them. All authentication mechanisms proposed act as mere gatekeepers, allowing users to enter. There are no checks after the initial authentication. Authorization decisions are based solely on the identity determined at the time of entry.

\subsection{VR application for telesurgery}
In todays connected world, it is possible for a surgeon sitting in one location and performing surgery on a patient in a different location. This is being done using robotic technology.

It is possible to see a time in the near future, where the surgeon performing telesurgery is using a VR system that provides the immersive experience of being in the operation theatre and performing the surgery in person. 

This problem presents numerous challenges. Firstly, we want to ensure that the surgery is indeed being performed by the surgeon. Secondly, it is necessary to ensure that there is continuous authentication to prevent unauthorized personal to take over the VR system from the physician. Thirdly, we want to ensure that patient confidentiality and HIPAA compliance are maintained. Business rules may require the doctor to be in certain location during surgery. 

\section{ADDITIONAL CONSIDERATIONS}\label{s4}
Security for virtual reality can be considered from the perspective of the fact that a virtual reality system is essentially a collection of peripherals such as cameras, microphones, motion trackers, screens, and controllers. However, the use of the technology can have implications that range from physical and privacy risks to ethical quandaries.

There are certain physical risk factors that exist for users of VR, such as risk of seizure, motion sickness, and physical discomfort. Concerns have also been reported involving vision damage or psychological harm due to usage \cite{1}. Due to the immersive nature of virtual reality, it is possible for malicious agents to intentionally trigger these symptoms. Interestingly, researchers have not only had success with this, but also with controlling the user’s movement. In an IEEE study, 87.5\% of subjects were able to have their movements controlled by the addition of supplementary content to their VR screen. For example, malicious software might add additional “objectives” or mechanics to games that induces the user to move in a particular direction. This has been dubbed the Human Joystick effect \cite{2}. A primary use of this technology is to trick a user to enter certain types of biometric authentication. Additionally, it can be used for physical attacks by inducing a user to move past the boundaries of their play area, potentially causing them to run into a wall. With high success rates for the current, less immersive implementations of VR, concerns should be had for future systems that may feature much more convincing virtual realities.
While convincing people to run into walls using VR might seem like a pointless pursuit, the technology that enables it is anything but benign. Attackers can use the relevant software to overlay images on existing software, making it seem like new elements have been added or removed. Researchers have been able to replicate this effect very consistently on the two most major VR systems, the HTC Vive and Oculus Rift by manipulation of files in the Steam software that they have in common. In doing so, they were able to not only implement overlays, but also set up independent sessions that allowed them to collect data on the user’s actions \cite{2}. Researchers have also examined and raised concerns about such technology being utilized in mixed reality (also known as augmented reality) which is an extension of virtual reality that allows the combination of virtual and real-world objects by means of either a partially clear VR screen or a video camera that has its feed combined with the necessary additions before being sent to the standard VR screen \cite{4}. Considering the vulnerabilities that have been established with existing systems, MR applications of these systems would also allow for more subtle attacks. Malicious agents might be able to deceive users by applying false images to screens, copying the user’s keystrokes or taps out of their peripheral vision, recording accurate observations on documents in the user’s field of view, or otherwise invading the users’ privacy. Though not all of these have been rigorously tested yet, researchers have been able to produce software that can reliably process videos from both AR cameras when the user uses their device from under 1.5 meters away. After this, the software was able to replicate observed 4-digit passwords with over 90\% success rate \cite{3}. 

Given this success rate, we can infer that human observation can produce better results yet. In addition, the most popular VR headsets, the HTC Vive and Oculus Rift, both feature mounted cameras and microphones that are constantly running, even if the device is inactive.

Another major concern when it comes to VR-related privacy lies in the social aspect of VR. In recent years, VR has become much more accessible to the common market, which has resulted in certain concerns regarding interactions between users in multiplayer settings and the more physically involved nature of the system. A 2019 study involving Virtual Reality Learning Environments, a type of immersive VR learning platform, showed that in addition to the known vulnerabilities of server-based interaction between multiple users such as denial-of-service attacks, further breaches could be induced resulting in the ability to mimic others via Elevation of Privilege or access confidential information \cite{5}. Due to the social nature of virtual reality, these types of threats need to be prevented more actively, as they can result in serious personal damages beyond what attackers are generally able to achieve in current social-interaction platforms such as social media. Given the advent of social VR software for entertainment and learning, the repercussions of essentially hijacking a person’s actions cannot be understated.

In addition to concerns regarding malicious entities, it is also important to analyze ethical concerns. VR developers and users have raised concerns about the ethical standards of companies that produce VR hardware and software. A prime example of this is Facebook’s ownership of Oculus, which has prompted to say that the company is “not afraid to manipulate… it’s all about the money to them ” \cite{1}. Another issue unique to VR ethical and privacy concerns is the widespread lack of permission requests involving data collection or usage of microphones and cameras. Many applications operate without any kind of privacy policy or privacy policies that are inadequately explained. Compared to the previously discussed privacy and security concerns, ethical concerns have a more attainable solution. The implementation of a common code of ethics and standards could change the nature of VR ethics. If developers and other entities were held to these standards by a large company like Valve Corporation or by the law, many of these concerns could be alleviated. The need for ethical codes is largely predicated by the personal nature of virtual reality- biometric and kinesiological reading of users is already a reality \cite{9}. Advertisers could take advantage of this to track and collect data on what users are looking at, for instance.

After adequately understanding the problems faced in the world of VR security, we should consider potential solutions. One major question is the process of authentication in VR technology. Because VR largely lacks the fine motor control necessary to simulate a keyboard, new methods must be created. Experiments have been performed with more simple methods such as mobile phone style PIN locks and slide locks, but these fall victim to aforementioned pitfalls in being easy to read and replicate. Even with only the motions of the controllers to base their choice on, most people were able to replicate the codes with a high degree of success \cite{11}. Fortunately, researchers have developed authentication systems with VR-specific mechanisms that offer additional security. 

The threat to a user’s security does not end after authentication is succeeded. We must also take care to implement additional security measures both to user inputs, outputs, interactions, and the software itself. An issue that arises with the nature of permission distributions is that virtual reality programs necessarily use a vast array of peripherals and their information. For example, even a simple augmented reality game will require access to motion sensors, controllers, cameras, and a microphone to function optimally. Solutions that have met some success involve anonymization of data used by the programs. Even if it is intercepted, it cannot be traced to the specific user. Unsurprisingly, data protection has also been successful when implemented \cite{4}. A solution that could work on a larger scale is input/output abstraction. This could be accomplished by the addition of an intermediary layer, for example, when Steam is used to boot its VR application. This would allow minimal access to the less secure program, meaning that even if it has vulnerabilities, it does not have information that could cause harm if lost.

Finally, it is worth noting the merits of properly implemented virtual reality. For example, the Swiss Federal Institute of Technology Lausanne has described a full security system monitoring physical and digital securities that uses VR to allow advanced multimedia control \cite{10}. Although the potential for both personal and large-scale security threats in virtual reality does exist, the potential benefits of proper utilization make investments in security advancements well worth the price. Virtual reality has the potent. 

\section{MULTIDEMNSIONAL AUTHENTICATION AND AUTHORIZATION }\label{s5}
In commercial applications of virtual reality, it is no longer sufficient to define that a given user has access to a certain resource. This decision needs to be nuanced, given that the authorization decision is no longer confined to actions within the confines of a computer. If we take the use case of an individual within the warehouse that is authorized to pick certain items of the shelf from certain areas of the warehouse but needs additional authorizations in other areas. This adds multiple dimensions to the identity of the user. Therefore, we need to know that basic identity has been determined. Then there is a need to know the location of the user, which places additional restrictions on the authorization. Thirdly, there may be a need for a different level of authentication involving biometrics or other such complexities to further access a third level of resources. In other words, the authorization decisions are not limited to the identity of the user but are also tied to environmental factors (like location, temperature, weather etc.) and to the type of authentication mechanism. 

It is therefore proposed that authentication is more complex than merely determining the identity of the individual. Each user can authenticate with different mechanisms, thus having a level of authentication. In addition, the authentication can incorporate other environmental factors like location, temperature etc. This multidimensional identity is then incorporated into the authorization process. This will provide a level of granularity in the security mechanisms where business rules and policies can be applied to ensure that the immersive real-time experience afforded within the VR context is not only secure based on the initial login, but also ensures that personal information and business intelligence are not compromised.

\section{Other ways to meet VR Security and Privacy}\label{s6}

\subsection{Blockchain applications to meet VR}
Today, VR has gone beyond the game, and people are ready to use this technology for art, tourism, and industrial purposes  \cite{v1, v2}. For this purpose, people need to pay a fee or get a ticket, and a blockchain is an excellent option for entirely secure payments  \cite{l1,a4}. As a distributed cryptographic ledger, a blockchain provides a platform that non-trusting parties can verifiably interact without a trusted authority \cite{choo2020blockchain,TAYLOR2020147,8917991}. There are different types of blockchain, private, public, and consortium, depending on how users log in and their access rights  \cite{a4,a5}. Blockchain-based VR often uses a general blockchain for commercial purposes so that people who want to use the platform can easily join it. The blockchain process in VR is such that all the information received from the user is first hashed, then the private key on the user's device is applied to it and generates a digital signature. Transaction data is sent to the peer and the digital signature on the peer-to-peer (P2P) network.
At this point, network members decrypt the device by applying the public key to the digital signature and comparing the resulting hash with the transaction data hash. In case of compliance, user authentication is performed \cite{a1,a2}. The consensus process is then chosen to determine the block's constructor to carry out the transaction, which results in the data being placed in the block and sent to the P2P network. Within the  P2P network, the transaction block is scanned, and if it is confirmed, the transaction is performed, and the data is recorded in the blockchain. As mentioned, this process has used several techniques, including hashing, asymmetric encryption, digital signature, etc., for authentication and secure data transfer. Therefore, the possibility of infiltration and abuse is very low. Also, by using smart contracts, people can be sure that they will receive the same services according to the agreement. The widespread use of this technology has raised concerns about the theft, hacking, dissemination, and copying of information, and as a result, has failed to limit their supply. Many companies are looking to find ways to use VR. A blockchain is a good option for creating decentralization, security, and transparency in sharing and using information. In addition, individuals can be assured that their personal data recorded in the virtual reality system will not be altered and that its uses can be easily traced. The security of using blockchain is technological advancement and creates a better user experience. We discuss some applications of blockchain to improve security and trust in virtual reality in the following.

The use of VR in e-commerce is also significant today. In this approach, people can review the products without a face-to-face consultation. And choose the ideal item. In this case, blockchain is used for payment security and to store product information and authorship. Therefore, illegal distribution and copyright infringement are prevented. For example, in the Cappasity platform \cite{12}, clothing retailers wanted to make an opportunity for customers to try out their designs in a virtual world rather than a physical one. In that case, they can protect their design rights using non-fungible tokens (NFT) \cite{n1}. As long as NFT is a unique and non-interchangeable unit of data stored on a blockchain, they can ensure their designs cannot be copied. This method is also widely used in virtual art exhibitions, and people can be sure that the products they are considering and buying are original. Another advantage of blockchain in VR is the management of information ownership and the allocation of access rights based on their role. For example, when VR is used to create a virtual classroom, only the teacher can change the educational content.

\subsection{Applying federated learning to meet VR}

Nowadays, machine learning (ML) is combined with various sciences to improve them and create new facilities \cite{a6,a7,9097894}. In VR, machine learning can also process people's data to create a more enjoyable user experience and add more interaction to VR. Furthermore, VR can decide and act better, more accurately, faster, and more intelligently with the device \cite{v3}. But it must be taken into account that the use of ML requires the collection and maintenance of information in a centralized manner. This feature threatens the privacy of information. On the other hand, if hackers attack this central server, they will access all information.

Federated learning (FL) can be an excellent option to solve this problem \cite{9153560,9521524,a9}. FL is an ML method that usually starts with a general model as a baseline. This general model is sent to the users; then, users train these local versions. In the second step, users send all the parameters of their learned model to the central server. Local servers then send their built-in model to the central server for aggregation. The central server collects the received models and sends the model updates to the nodes. This process is repeated until the desired level of accuracy is reached. As long as FL sends only hyper-parameters instead of sending all data to the central server, it can increase security and privacy \cite{a9}. FL can prevent security attacks and misuse of information by providing integrated access to information while protecting their privacy \cite{9424138,a9,a8,9631164}. For example, people can use their mobile phones to check and test their home decoration virtual reality. In this case, there is no need to send the people's primary information, like all the details of the people's home, to the centralized server. Instead, this information creates a local model on a person's mobile phone and sends the hyper-parameter update to the central server to complete the public model. Furthermore, as we can see, FL assists in securing ML-based VR for users and industry owners. Moreover, although local models learn from the global model, they create a unique model because they use the features and personal data in the device. In other words, instead of personalization on a central server, personalization is done on users' devices according to their preferences and choices. This feature turns VR into a secure and intelligent tool that can learn from the user and adjust the service. For instance, it can adjust the image quality for people who have eye problems. In another type of personalization, if a person uses VR to visit an art exhibition, the VR recommendation system directs them to previous popular works and styles they have already purchased. In the traditional machine learning model, it was necessary to collect all the information of users' previous purchases and store it in the central server for these intelligent suggestions. In other words, personalization was done at the cost of losing privacy. However, in the  FL approach, both goals can be achieved simultaneously. 

\section{CONCLUSION}\label{s7}
The challenges posed by VR are a cause to expand the authentication and authorization processes.  Primary factors that contribute to this are:
\begin{itemize}
   	\item {	VR is headed in a direction where it is no longer tethered to a PC and is a self-contained wearable device.}
	\item {	VR headsets provide immersive experience by cutting of the view of the surrounding, making it impossible to be aware of predatory observers.}
	\item{ VR experience involves performing physical tasks in various environments and must be constrained based on the physical world that is different from the virtual world that the user is immersed in.}
\end{itemize}

Literature shows that various authentication mechanisms have been proposed to address the issues of being observed to compromise the identity. Some of these mechanisms leverage biometrics and other more advanced methodologies. The complexity of the authentication mechanisms adds additional cost to the VR infrastructure.  From a practical perspective, adding extra hardware to the VR headset is going to be a tradeoff with the level of comfort for the user, especially when there is prolonged use.

Commercial applications typically have different classes of users. All of these users may not require all levels of authentication. Therefore, cost of the VR technology may be contained in organizations by providing different levels of hardware. 

There is however a definite need for granularity in the security decision to justify a multi-dimensional authentication and authorization mechanism.

\bibliographystyle{ieeetr}
\bibliography{rf}
\end{document}